\documentclass[showpacs,aps,graphicx]{revtex4}
\usepackage{graphicx}

\begin{document}

\title{Multiparty quantum secret sharing with pure entangled states and decoy photons\footnote{Published in \emph{Physica} A 381 (2007) 164-169.}}

\author{ Ping Zhou$^{1,2,3}$, Xi-Han Li$^{1,2,3}$, Yu-Jie Liang$^{1,2,3}$, Fu-Guo Deng$^{1,2,3}$\footnote{
Email address: fgdeng@bnu.edu.cn}, Hong-Yu Zhou$^{1,2,3}$ }
\address{$^1$ Key Laboratory of Beam Technology and Material
Modification of Ministry of Education, Beijing Normal University,
Beijing 100875, P. R. China\\
$^2$ Institute of Low Energy Nuclear Physics, and Department of
Material Science and Engineering, Beijing Normal University, Beijing
100875, P. R. China\\
$^3$ Beijing Radiation Center, Beijing 100875, P. R. China}
\date{\today }

\begin{abstract}
We present a scheme for multiparty quantum  secret sharing of a
private key with pure entangled states and decoy photons. The boss,
say Alice uses the decoy photons, which are randomly in one of the
four nonorthogonal single-photon states, to prevent a potentially
dishonest agent from eavesdropping freely. This scheme  requires the
parties of communication to have neither an ideal single-photon
quantum source nor a maximally entangled one, which makes this
scheme more convenient than others in a practical application.
Moreover, it has the advantage of having high intrinsic efficiency
for qubits and exchanging less classical information in principle.
\bigskip

\emph{Keywords:} Quantum secret sharing; Quantum communication; Pure
entangled states; Decoy photons.

\end{abstract}
\pacs{03.67.Hk, 03.65.Ud} \maketitle

The principles in quantum mechanics, such as the uncertain relation,
the correlation of entangled quantum systems and the collapse in
quantum measurement, provide some novel ways for secure
communication. For instance, quantum key distribution (QKD), whose
goal is used to create a private key between two authorized users,
has become one of the most mature applications of quantum
information techniques
\cite{Gisin,BB84,Ekert91,BBM92,ABC,CORE,BidQKD,LongLiu}. The
noncloning theorem forbids a vicious eavesdropper, say Eve, to
eavesdrop the quantum information transmitted through a quantum
channel freely. The action of Eve will inevitably disturb the
quantum systems transmitted and leave a trace in the outcomes
obtained by the receiver. The two authorized parties can find out
Eve's eavesdropping by comparing a subset of the outcomes in public.
This principle has been also applied to other branches of quantum
communication, such as quantum secure direct
communication\cite{two-step,QOTP,Wangc}, deterministic quantum
communication \cite{decoy3,zhangzj,zhangsPRA,lixhjp},  quantum
secret report \cite{dengreport}, quantum broadcast
communication\cite{wangbroadcast}, quantum secret conference
\cite{qsc}, quantum dialogue \cite{dialogue}, and so on.

Quantum secret sharing (QSS) is the quantum counterpart of the
classical secret sharing. In a secret sharing, a boss, say Alice,
has two agents (Bob and Charlie) who are at a remote place, and she
wants to send her instruction to her agents for dealing with her
business. However Alice suspects that one of her agents may be
dishonest, and she does not know who is the dishonest one. Alice
believes that the honest one can keep the potentially dishonest from
doing harm to her benefits if they both coexist in the process of
the business. For the security of the secret message, say $M_A$,
Alice will divide it into two pieces, $M_B$ and $M_C$, and then
sends them to Bob and Charlie, respectively. If and only if Bob and
Charlie cooperate, they can read out the message $M_A=M_B \oplus
M_C$; otherwise none can obtain a useful information about the
secret message. As classical signal is in the eigenvectors of a
quantum operator, it can be copied freely and fully. That is to say,
it is impossible in principle to share a secret message with
classical physics. When quantum mechanics enters the field of
information, the story is changed. In 1999, Hillery, Bu\v{z}ek and
Berthiaume (HBB99) \cite{HBB99} proposed an original QSS scheme for
creating a private key among three parties with a three-particle
Greenberger-Horne-Zeilinger (GHZ) state $\vert
GHZ\rangle=\frac{1}{\sqrt{2}}(\vert 000\rangle +\vert 111\rangle)$,
the maximally entangled three-particle state. Here $\vert 0\rangle$
and $\vert 1\rangle$ are the two eigenvectors of the measuring basis
(MB) $Z$ (for example the z-direction of $1/2$-spin). Now, there are
a great number of QSS schemes, such as the schemes
\cite{HBB99,KKI,longQSS,Karimipour,Bandyopadhyay,cpyang,COREQSS,Guo,qssjpa,yanpra,Zhang,improving,zhangpa,chenp}
for creating a private key and those
\cite{Peng,dengQSTS,QSTS2,QSTS3} for sharing an unknown quantum
state.

Almost all the existing QSS schemes are based on either maximally
entangled quantum signals
\cite{HBB99,KKI,longQSS,Karimipour,Bandyopadhyay,cpyang,COREQSS,zhangpa,chenp}
or an ideal single-photon quantum signal
\cite{Guo,qssjpa,yanpra,Zhang,improving}, which makes them difficult
in a practical application. On one hand, a practical ideal
single-photon source cannot be obtained with present techniques
although people can in principle produce a single photon. On the
other hand, an entangled source usually generates a pure entangled
state because of the property of asymmetry in the source.

In this paper, we will present a scheme for quantum secret sharing
with pure entangled states, not maximally entangled ones. The boss
Alice exploits some decoy photons to forbid a potentially dishonest
agent to steal the information about the private key obtained by
another agent. This scheme has the advantage of having high
intrinsic efficiency for qubits and exchanging little classical
information. Moreover, it does not require the parties of
communication to have an ideal single-photon quantum source, which
is not available in a practical application with the present
techniques, or a maximally entangled quantum source. Thus this QSS
scheme is more convenient than others.

Now, let us describe the principle of our QSS scheme. For
simplicity, we first describe it  with two agents, i.e., Bob and
Charlie, and then generalize it to the case with $N$ agents. For the
case with two agents, Alice should first prepare a sequence of pure
entangled photon pairs $S$. Each  pair is in one of the four states
$\{\vert \phi\rangle_{BC}, \vert \phi'\rangle_{BC}, \vert
\psi\rangle_{BC},\vert \psi'\rangle_{BC}\}$.
\begin{eqnarray}
\vert \phi\rangle_{BC}&=&(\alpha \vert 00\rangle + \beta \vert
11\rangle)_{BC},\\
\vert \phi'\rangle_{BC}&=&(\alpha \vert 11\rangle + \beta \vert
00\rangle)_{BC},\\
\vert \psi\rangle_{BC}&=&(\alpha \vert 01\rangle + \beta \vert
10\rangle)_{BC},\\
\vert \psi'\rangle_{BC}&=&(\alpha \vert 10\rangle + \beta \vert
01\rangle)_{BC},
\end{eqnarray}
where
\begin{eqnarray}
\vert \alpha\vert^2 + \vert \beta \vert^2 =1.
\end{eqnarray}
Suppose that Alice's entangled source produces an entangled pair in
the state $\vert \phi\rangle_{BC}=(\alpha \vert 00\rangle + \beta
\vert 11\rangle)_{BC}$ in each signal interval, Alice can obtain the
entangled pair sequence $S$ by operating some of the pairs with the
two unitary operations $U_{0}=\left\vert 0\right\rangle \left\langle
0\right\vert +\left\vert 1\right\rangle \left\langle 1\right\vert$
and $U_{1}=\left\vert 1\right\rangle \left\langle 0\right\vert
+\left\vert 0\right\rangle \left\langle 1\right\vert$, i.e.,
\begin{eqnarray}
\vert \phi'\rangle_{BC}=(U_1^B \otimes U_1^C)\vert
\phi\rangle_{BC},\\
\vert \psi\rangle_{BC}=(U_0^B \otimes U_1^C)\vert
\phi\rangle_{BC},\\
\vert \psi'\rangle_{BC}=(U_1^B \otimes U_0^C)\vert \phi\rangle_{BC}.
\end{eqnarray}
Alice divides the sequence $S$ into two sequences $S_B$ and $S_C$.
The sequence $S_B$ is made up of all the $B$ photons of   the photon
pairs in the sequence $S$. All the $C$ photons compose of the
partner particle sequence $S_C$, similar to Refs.
\cite{LongLiu,two-step,QOTP,Wangc,decoy3,lixhjp}. Different from the
Karlsson-Koashi-Imoto (KKI) QSS scheme \cite{KKI}, the quantum
information carries in this scheme are a sequence of pure entangled
stats. The photon $B$ and the photon $C$ in a pure entangled state
are completely correlated when they are measured with the MB $Z$,
but not with the MB $X=\{\vert \pm x\rangle
=\frac{1}{\sqrt{2}}(\vert 0\rangle \pm \vert 1\rangle)\}$. For
instance,
\begin{eqnarray}
\vert \phi\rangle_{BC}&=&(\alpha \vert 00\rangle + \beta \vert
11\rangle)_{BC} =\frac{1}{2}[(\alpha + \beta)(\vert +x\rangle\vert
+x\rangle +
\vert -x\rangle\vert -x\rangle)\nonumber\\
&&\,\, + (\alpha - \beta)(\vert +x\rangle\vert -x\rangle + \vert
-x\rangle\vert +x\rangle)].
\end{eqnarray}
That is, on one hand, the security of the quantum secret sharing
with pure entangled states is lower than that with Bell states if
the parties use the two MBs $Z$ and $X$ to measure their photon
pairs for the eavesdropping check directly \cite{decoy3}. On the
other hand, the quantum source is more convenient than maximally
entangled states \cite{decoy3} as the asymmetry in a practical
quantum source makes the photon pairs in a pure entangled state, not
a maximally one.

For ensuring the security of the transmission of the photon
sequences $S_B$ and $S_C$, Alice should add some decoy photons in
these two sequences before she sends $S_B$ and $S_C$ to Bob and
Charlie, respectively. The decoy photon technique was proposed first
by Li \emph{et al.} \cite{decoy,decoy2} in QKD network, and now it
has been applied to other branches of quantum communication, such as
deterministic secure quantum communication \cite{decoy3}, quantum
secret report \cite{dengreport} and quantum secret conference
\cite{qsc}. The principle of the decoy photon technique is that
Alice prepares some photons which are randomly in one of the four
nonorthogonal states $\{\vert 0\rangle, \vert 1\rangle, \vert
+x\rangle, \vert -x\rangle\}$ and then inserts them into the
sequences $S_B$ and $S_C$. As the states and the positions of the
decoy photons are unknown for all the parties of the communication
except for Alice herself, the eavesdropping done by an eavesdropper
will inevitably disturb these decoy photons and will be detected,
similar to Bennett-Brassard 1984 (BB84) protocol \cite{BB84} and its
modified version \cite{ABC}. The number of the decoy photons is not
required to be very large, just large enough for checking
eavesdropping. Still, it is unnecessary for Alice to prepare her
decoy photons with an ideal single-photon source. She can get them
by measuring the photon $B$ in a pure entangled state $\vert
\phi\rangle_{BC}$ and manipulating the other photon $C$ with some
unitary operations. For example, if Alice wants to make her decoy
photon in the state $\vert +x\rangle$, she measures the photon $B$
in the pure entangled state $\vert \phi\rangle_{BC}$ with the MB $Z$
and then performs a Hadamard ($H$) operation on the photon $C$ when
she obtains the outcome $\vert 0\rangle_B$, otherwise she performs
the operation $H\otimes U_1$ on the  photon $C$. Here
$H=(1/\sqrt{2})(\vert 0\rangle\langle 0\vert + \vert 1\rangle\langle
0\vert + \vert 0\rangle\langle 1\vert - \vert 1\rangle\langle
1\vert)$. As the analysis of the decoy photons is as same as that in
the BB84 QKD protocol \cite{BB84}, our QSS scheme is secure if Alice
exploits her decoy photons to forbid the potentially dishonest agent
to eavesdrop freely.

Suppose that Alice codes the states $\{\vert \phi\rangle_{BC}, \vert
\phi'\rangle\}_{BC}$ as $0$ and codes the states $\{\vert
\psi\rangle_{BC}, \vert \psi'\rangle_{BC}\}$ as 1. Our three-party
QSS scheme with pure entangled states can work with following steps.

(S1) Alice prepares a sequence of pure entangled two-photon states
$S$,  $N$ ordered photon pairs. Each photon pair $BC$ is randomly in
one of the four states $\{\vert \phi \rangle_{BC}, \vert
\phi'\rangle_{BC}, \vert \psi\rangle_{BC}, \vert
\psi'\rangle_{BC}\}$. She divides the sequence $S$ into two partner
particle sequences $S_B$ and $S_C$. The sequence $S_B$ ($S_C$) is
made up of the photons $B$ ($C$) in the ordered $N$ photon pairs.
Alice prepares $2k$ ($k<<N$) decoy photons by measuring the photons
$B$ in some photon pairs $BC$ and operating the remaining photons
$C$ with the two unitary operations $U_i$ ($i=0,1$) and the $H$
operation. She inserts randomly $k$ decoy photons into the sequence
$S_B$ and the other $k$ decoy photons into the sequence $S_C$.

(S2) Alice sends the sequence $S_B$ and $S_C$ to Bob and Charlie,
respectively.

(S3) Bob and Charlie measure their photons in the sequences $S_B$
and $S_C$ independently with the two MBs $Z$ and $X$.

If Bob and Charlie have the capability of storing their quantum
states, they can measure their photons in the sequences $S_B$ and
$S_C$ in the same way as those in Refs. \cite{two-step,QOTP}. That
is, Alice first tells Bob and Charlie which are the decoy photons
and their states, and then Bob and Charlie measure the decoy photons
with the same MBs as those used by Alice for preparing them. For the
other photons, Bob and Charlie both choose the MB $Z$ to measure
them. In this way, all the decoy photons can be used for checking
eavesdropping. The $k$ decoy photons in the sequence $S_B$ ($S_C$)
can be used to check the security of Alice-Bob (Alice-Charlie)
quantum line in the same way as the BB84 QKD protocol \cite{BB84}
which has been proven to be secure for generating a private key
\cite{BB84security}.

Without quantum memory, Bob (Charlie) can measure the photons in the
sequence $S_B$ ($S_C$) in the same way as that used in the modified
BB84 QKD protocol \cite{ABC}. That is to say, Bob (Charlie) measures
his photons by using the MB $X$ with the probability $p$ ($p<<1/2$
if $N$ is large enough) and  the MB $Z$ with the probability $1-p$.
In this way, half the decoy photons measured by Bob (Charlie) in the
sequence $S_B$ ($S_C$) are useful for checking eavesdropping as the
MBs for preparing and measuring them are the same ones. In this
time, Alice and Bob (Alice and Charlie) can analyze the error rate
of the decoy photons in the sequence $S_B$ ($S_C$) with the refined
analysis technique discussed in the modified BB84 QKD protocol
\cite{ABC}. In detail, they divide the useful decoy photons into two
groups, the one they both choose the MB $Z$ and the other one they
both choose the MB $X$. They analyze the error rates of these two
groups independently. When the error rates of the two groups of
decoy photons both are lower than the threshold $\eta_t$, Alice
believes that the transmission between her and her agents is secure
or the information leaked to a potentially dishonest agent is
negligible.

(S4) If Alice confirms that the two quantum lines, i.e., Alice-Bob
and Alice-Charlie, both are secure, Alice and her agents distill a
private key $K_A=K_B \oplus K_C$ with the other outcomes they all
choose the MB $Z$, similar to QKD \cite{Gisin}; otherwise they
repeat their QSS from the beginning. Here $K_A$, $K_B$ and $K_C$ are
the key obtained by Alice, Bob and Charlie, respectively.

As pointed out in Ref. \cite{KKI}, a QSS scheme is secure if it can
prevent the potentially dishonest agent from eavesdropping freely.
In this QSS scheme, Alice exploits her decoy photons to forbid her
agents to eavesdrop freely. As the states and the positions of the
decoy photons are unknown for Bob and Charlie, either Bob or Charlie
will be detected if he wants to steal the information about the key
obtained by the other agent. Thus this QSS scheme can be made to be
secure. On the other hand, as the agents choose the MB $Z$ with a
lager probability $1-p$ to measure their photons used for creating
the private key, the intrinsic efficiency of qubits $\eta_q \equiv
\frac{q_u}{q_t}$ is far larger than that in KKI QSS scheme
\cite{KKI} as p is much smaller than $1/2$. Here $q_u$ is the number
of the useful qubits and $q_t$ is that of the total qubits
transmitted.

It is straightforward to generalize this scheme to the case with $M$
agents, say Bob$_i$ ($i=1,2,\cdots, M$). Similar to the case with
two agents, Alice prepares a sequence of pure entangled $M$-photon
quantum systems $S'$  (ordered $N$ pure entangled quantum systems),
and each quantum system is randomly in one of the four states
$\{\vert \Phi\rangle, \vert \Phi'\rangle, \vert \Psi\rangle, \vert
\Psi'\rangle\}$. Here
\begin{eqnarray}
\vert \Phi\rangle_{B_1B_2\cdots B_M} &=& (\alpha \vert 00\cdots
0\rangle + \beta \vert
11\cdots 1\rangle)_{B_1B_2\cdots B_M}, \\
\vert \Phi'\rangle_{B_1B_2\cdots B_M} &=& (\alpha \vert 11\cdots
1\rangle + \beta \vert
00\cdots 0\rangle)_{B_1B_2\cdots B_M}, \\
\vert \Psi\rangle_{B_1B_2\cdots B_M} &=& (\alpha \vert 00\cdots
1\rangle + \beta \vert 11\cdots 0\rangle)_{B_1B_2\cdots
B_M}, \\
\vert \Psi'\rangle_{B_1B_2\cdots B_M} &=&(\alpha \vert 11\cdots
0\rangle + \beta \vert 00\cdots 1\rangle)_{B_1B_2\cdots B_M}.
\end{eqnarray}
Alice divides the sequence $S'$ into $M$ partner photon sequences
$S'_{B_1}$, $S'_{B_2}$, $\cdots$, and $S'_{B_M}$. The sequence of
$S'_i$ ($i=1,2,\cdots, M$) is made up of the photons $B_i$ in all
the ordered quantum systems. Also Alice insets randomly $k$ decoy
photons, which are prepared by measuring a photon in a pure
entangled state and operating the remaining photons with unitary
operations $U_0$, $U_1$ and $H$, into each partner photon sequence
$S_i$ before she sends it to the agent Bob$_i$. If all the agents
have the capability of storing their quantum states, Alice first
tells the agents which are the decoy photons and their states in the
partner photon sequences when all the agents have received their
sequences, and then the agents measure the decoy photons with the
same MBs as those used by Alice for preparing them and measure all
the pure entangled states with the MB $Z$. Without quantum memory,
the agents should measure their photons in the same way as the case
with two agents. That is, all the agents measure their photons by
using the MB $X$ with a small probability $p$ and the MB $Z$ with
the probability $1-p$. In this time, the rate of the useful qubits
to all those transmitted is $p_u=(1-p)^M$. As the same as the case
with two agents, the security of each of the partner photons $S_i$
is ensured by the decoy photons, the same as the BB84 QKD protocol
\cite{BB84} or its modified version \cite{ABC}. In other words, this
QSS scheme with $N$ agents can be made to be secure.

As proven by Deng \emph{et al.} \cite{proving} that the one-way QSS
schemes based on entanglement and a collective eavesdropping check,
such as the two famous QSS schemes, the HBB99 QSS scheme and the KKI
QSS scheme, are insecure with a lossy quantum channel if the parties
only exploit the correlation of the entangled quantum systems to
check eavesdropping, this QSS scheme is an optimal one. It has the
following advantages obviously: (a) The quantum signals are a
sequence of pure entangled states, not maximally entangled ones,
which makes it more convenient than others in a practical
application. (b) The boss Alice exploits some decoy photons to
ensure the security of this scheme, which is just the requirement of
a secure one-way QSS scheme based on entanglement and a collective
eavesdropping check in a practical application \cite{proving}. (c)
It requires the boss has neither an ideal single-photon quantum
signal source nor a maximally entangled source. (d) It does not
requires the parties exchange a large number of classical
information as the agents choose a large probability to measure
their photons with the MB $Z$. (e) Its intrinsic efficiency for
qubits is very high, approaching 100\% when the number of bits in
the private key $K_A$ is large enough.

In conclusion, we have presented a QSS scheme with pure entangled
states and decoy photons. As this scheme requires the parties to
have neither an ideal single-photon quantum source nor a maximally
entangle one, it is more convenient in a practical application than
others. Whether the agents have quantum memory or not, this scheme
has the advantage of having a high intrinsic efficiency for qubits
and exchanging little classical information. For a one-way QSS based
on entanglement and a collective eavesdropping check, it is useful
for the boss Alice to make at least the qubits used for checking
eavesdropping in single-particle states, not entangled ones
\cite{proving}. Thus this scheme with decoy photons is an optimal
one.

\bigskip

This work is supported by the National Natural Science Foundation of
China under Grant Nos. 10604008 and 10435020, and Beijing Education
Committee under Grant No. XK100270454.


\begin{thebibliography}{99}

\bibitem{Gisin} N. Gisin, G. Ribordy, W . Tittel,  H. Zbinden,
Rev. Mod. Phys.74 (2002) 145.


\bibitem{BB84} C. H.  Bennett, G. Brassard, in: Proceeding of the IEEE International Conference on Computers, Systems
and Signal Processing, Bangalore, India, IEEE, New York, 1984, P.
175.



\bibitem{Ekert91} A. K. Ekert,  Phys. Rev. Lett. 67
(1991)  661.

\bibitem{BBM92} C. H. Bennett, G. Brassard, N. D. Mermin,
 Phys. Rev. Lett. 68 (1992)  557.




\bibitem{CORE} F. G. Deng, G. L. Long, Phys. Rev. A 68 (2003)  042315.

\bibitem{BidQKD} F. G. Deng, G. L. Long, Phys. Rev. A 70 (2004)
012311.


\bibitem{LongLiu} G. L. Long, X. S. Liu, Phys. Rev. A 65 (2002) 032302.


\bibitem{ABC} H. K.  Lo, H. F. Chau, M. Ardehali
 J. Cryptology 18 (2005)  122.


\bibitem{two-step} F. G. Deng, G. L. Long, X. S. Liu, Phys. Rev. A
68 (2003) 042317.

\bibitem{QOTP} F. G. Deng, G. L. Long,  Phys. Rev. A 69 (2004)  052319.

\bibitem{Wangc} C. Wang, F. G. Deng, Y. S. Li, X. S. Liu, G. L. Long, Phys. Rev. A
71 (2005) 044305;
 C. Wang, F. G. Deng, G. L. Long, Opt. Commun. 253 (2005)  15.


\bibitem{decoy3} X. H. Li, F. G. Deng, C. Y. Li, Y. J. Liang, P. Zhou, H. Y. Zhou, J. Korean Phys. Soc. 49
(2006)  1354.

\bibitem{lixhjp} X. H. Li, F. G. Deng, H. Y. Zhou,  Phys. Rev. A 74 (2006)  054302.

\bibitem{zhangzj} Z. X. Man, Z. J. Zhang,  Y. Li, Chin.
Phys. Lett. 22 (2005)   18;
 F. L. Yan,  X. Zhang,  Euro. Phys. J. B 41 (2004)    75;
 T. Gao, F. L. Yan,  Z. X. Wang,  Chin. Phys. 14 (2005)  893.






\bibitem{zhangsPRA} Q. Y.  Cai, B. W. Li Chin. Phys. Lett.
 21  (2004)  601; A. D. Zhu, Y. Xia, Q. B. Fan, S. Zhang,
 Phys. Rev. A  73 (2006)  022338;
 J. Wang, Q. Zhang, C. J. Tang, Phys. Lett. A
 358   (2006) 256;
 H. J. Cao,  H. S. Song,  Chin. Phys. Lett. 23 (2006)  290.






\bibitem{dengreport} F. G. Deng, X. H. Li, C. Y. Li, P. Zhou, Y. J. Liang, H. Y. Zhou, Chin. Phys. Lett. 23
(2006)  1676.


\bibitem{wangbroadcast} J. Wang, Q. Zhang, C. J. Tang,
quant-ph/060179.




\bibitem{qsc} X. H. Li, C. Y. Li, F. G. Deng, P. Zhou, Y. J. Liang, H. Y. Zhou,  Chin. Phys. Lett. 24
(2007)  23.


\bibitem{dialogue} B. A. Nguyen,  Phys. Lett. A 328
(2004)   6.




\bibitem{HBB99} M. Hillery, V. Bu\v{z}ek, A. Berthiaume,  Phys. Rev. A 59
(1999)  1829.

\bibitem{KKI} A. Karlsson, M. Koashi, N. Imoto, Phys. Rev. A 59
(1999)  162.

\bibitem{longQSS}L. Xiao, G. L. Long, F. G. Deng, J. W. Pan,  Phys. Rev. A
69 (2004) 052307; F. G. Deng, P. Zhou, X. H. Li, C. Y. Li, H. Y.
Zhou,  Chin. Phys. Lett. 23 (2006)  1084.




\bibitem{Karimipour} V. Karimipour, A. Bahraminasab,   S. Bagherinezhad, Phys.
Rev. A  65 (2002)  042320.


\bibitem{cpyang} C. P. Yang, J. Gea-Banacloche, J. Opt. B: Qantum
Semiclass. Opt.  3 (2001)  407.

\bibitem{Bandyopadhyay} S. Bandyopadhyay, Phys. Rev. A 62  (2000) 012308.



\bibitem{COREQSS}F. G. Deng, G. L. Long, H. Y. Zhou, Phys. Lett. A 340 (2005)   43;  F. G. Deng, X. H. Li, C. Y. Li, P. Zhou,
 H. Y. Zhou,   Phys. Lett. A 354 (2005)   190.


\bibitem{chenp} P Chen, F. G. Deng, G. L. Long,  Chin. Phys. 15
(2006) 2228.



\bibitem{zhangpa} Z. J. Zhang, Physica A 361 (2006) 233.


\bibitem{Guo} G. P. Guo,  G. C. Guo,  Phys. Lett. A 310
(2003)  247.

\bibitem{qssjpa} F. G. Deng, H. Y. Zhou, G. L. Long,  Phys. Lett. A 337
(2005)  329; F. G. Deng, H. Y. Zhou, G. L. Long,  J. Phys. A 39
(2006) 14089.



\bibitem{yanpra} F. L. Yan, T. Gao  Phys. Rev. A 72
(2005)  012304.

\bibitem{Zhang} Z. J. Zhang, Y. Li, Z. X. Man,  Phys. Rev. A 71 (2005)
044301.

\bibitem{improving} F. G. Deng, X. H. Li, H. Y. Zhou, Z. J. Zhang,  Phys. Rev.
A  72  (2005) 044302.





\bibitem{Peng} Y. M. Li, K. S. Zhang, K. C. Peng,  Phys. Lett. A 324
(2004)  420.


\bibitem{dengQSTS} F. G. Deng, C. Y. Li, Y. S. Li, H. Y. Zhou, Y. Wang, Phys. Rev. A
72 (2005) 044301;  F. G. Deng, X. H. Li, C. Y. Li, P. Zhou, H. Y.
Zhou,  Phys. Rev. A 72 (2005)  022338.

\bibitem{QSTS2} X. H. Li, P. Zhou, C. Y. LI, H. Y. Zhou, F. G. Deng,  J. Phys. B
39 (2006) 1975;
 F. G. Deng, X. H. Li, C. Y. Li, P. Zhou, H. Y. Zhou, Eur. Phys. J. D 39 (2006)  459.


\bibitem{QSTS3}Z. J. Zhang, Eur. Phys. J. D 33 (2005)
133;  Y. Q. Zhang, X. R.  Jin,  S. Zhang, Chin. Phys. 15
 (2006)  2252.




\bibitem{decoy} C. Y. Li, H. Y. Zhou, Y. Wang, F. G. Deng,   Chin. Phys. Lett. 22 (2005)
1049.


\bibitem{decoy2} C. Y. Li, X. H. Li, F. G. Deng, P. Zhou, Y. J. Liang, H. Y. Zhou,  Chin. Phys. Lett. 23
(2006) 2896.


\bibitem{BB84security} H. K. Lo,  H. F. Zhau,  Science 283 (1999)  2050; P. W. Shor, J. Preskill,  Phys.
Rev. Lett. 85 (2000)  441.


\bibitem{proving} F. G.  Deng, X. H. Li, H. Y. Zhou, arXiv:
0705.0279.

\end{thebibliography}
\end{document}